\documentclass{article}

\usepackage[left=1in,right=1in,top=1in,bottom=1in]{geometry} %
\usepackage{listings} %
\usepackage{sectsty} %
\allsectionsfont{\sffamily} %
\usepackage{authblk} %
\usepackage{graphicx} %
\usepackage{float} %
\usepackage{hyperref} %
\usepackage{url} %
\usepackage{gensymb} %
\usepackage[backend=biber,style=nature,sorting=none]{biblatex} %
\usepackage{abstract}
\providecommand{\keywords}[1]{\textbf{\textit{Keywords --}} #1}

\begin{document}

\title{\sffamily A Raspberry Pi-based, RFID-equipped birdfeeder for the \\remote monitoring of wild bird populations\rmfamily}

\date{}

\author[a,b,1]{\normalsize Mason Youngblood}

\affil[a]{\scriptsize Department of Psychology, The Graduate Center, City University of New York, New York, NY, USA}
\affil[b]{\scriptsize Department of Biology, Queens College, City University of New York, Flushing, NY, USA\newline\textsuperscript{1}myoungblood@gradcenter.cuny.edu}

\maketitle

\begin{abstract}
	\vspace*{-10pt}
	
	Radio-frequency identification (RFID) is an increasingly popular wireless technology that allows researchers to monitor wild bird populations from fixed locations in the field. Our lab has developed an RFID-equipped birdfeeder based on the Raspberry Pi Zero W, a low-cost single-board computer, that collects continuous visitation data from birds tagged with passive integrated transponder (PIT) tags. Each birdfeeder has a perch antenna connected to an RFID reader board on a Raspberry Pi powered by a portable battery. When a tagged bird lands on the perch to eat from the feeder, its unique code is stored with the date and time on the Raspberry Pi. These birdfeeders require only basic soldering and coding skills to assemble, and can be easily outfitted with additional hardware like video cameras and microphones. We outline the process of assembling the hardware and setting up the operating system for the birdfeeders. Then, we describe an example implementation of the birdfeeders to track house finches (\textit{Haemorhous mexicanus}) on the campus of Queens College in New York City.
	
	\\\\
	
\end{abstract}

\section*{Introduction}

Radio frequency identification (RFID) is an increasingly popular wireless technology that allows researchers to monitor wild bird populations at fixed locations in the field \cite{Bonter2011,Ponchon2013}. Individual birds are tagged with passive integrated transponder (PIT) tags with small microchips that transmit a unique code when activated by close proximity to an antenna connected to an RFID reader \cite{Ousterhout2014}. The lack of an internal power source means that these tags are small enough to have negligible effects on survival \cite{Adelman2015}. In addition, birds with PIT tags do not have to be recaptured and continuously handled to collect data, making this method much less invasive than traditional capture-recapture.

In the past, the high cost of commercial RFID readers has been a financial barrier to their use in research. Previous DIY readers are significantly cheaper, but typically require a background in circuit-building \cite{Bridge2016,Ibarra2015,Zarybnicka2016}. More recently, researchers have begun to implement more user-friendly DIY readers based on hobbyist electronics \cite{Bridge2019}. Our lab has developed an RFID-equipped birdfeeder based on the Raspberry Pi Zero W, a low-cost single-board computer with wireless LAN. Each birdfeeder has a perch antenna connected to an RFID reader board on a Raspberry Pi powered by a portable battery. When a tagged bird lands on the perch to eat from the feeder, its unique code is stored with the date and time on the Raspberry Pi. All collected data is then backed-up to the cloud using a personal hotspot. These birdfeeders require only basic soldering and coding skills to assemble, and can be easily outfitted with additional hardware like video cameras and microphones.

Below, we outline the process of assembling the hardware and setting up the operating system for the birdfeeders. Then, we describe an example implementation of the birdfeeders to track house finches (\textit{Haemorhous mexicanus}) on the campus of Queens College in New York City, and highlight several minor technical issues that we encountered using this system.

\begin{figure}[H]
\centering
\includegraphics[width=0.35\textwidth]{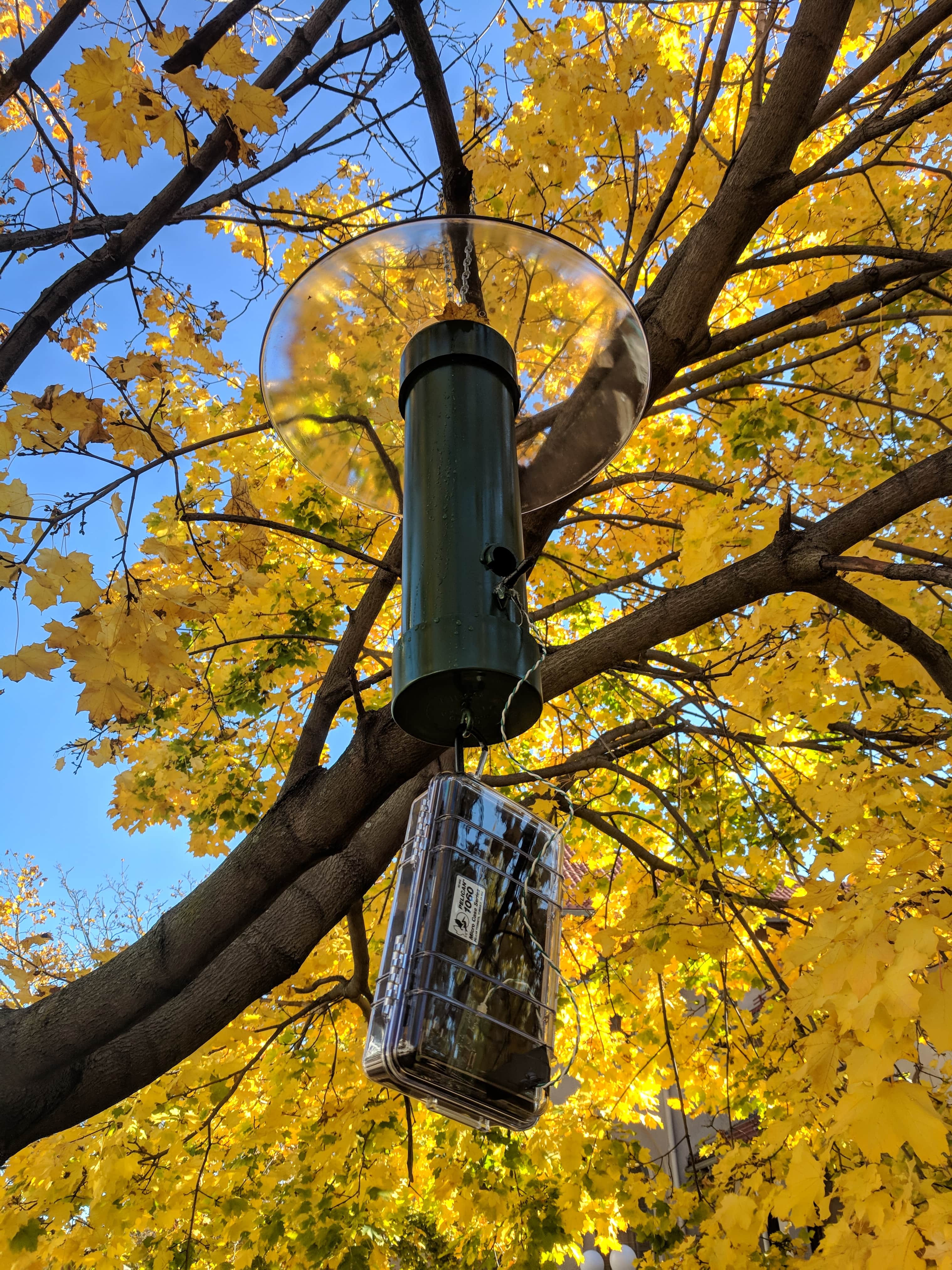}
\caption{One of the RFID-equipped birdfeeders hanging in a tree on the Queens College campus.}
\label{birdfeeder}
\end{figure}

\section*{Hardware}

\subsection{Raspberry Pi}

The Raspberry Pi Zero W\footnote{\url{https://www.raspberrypi.org/products/raspberry-pi-zero-w/}} is an ideal single-board computer for remote monitoring setups because of its low power consumption and built-in wireless LAN. Users who do not plan on backing up data to the cloud can choose to use the Raspberry Pi Zero instead. Each computer will require a microSD card for the operating system and internal storage. We also recommend that you purchase a mini-HDMI to HDMI adaptor and micro-USB to USB adaptor to setup the system with a monitor and keyboard. Before the computer can be used with the RFID reader, a standard 40-pin GPIO header needs to be soldered to the board.

\subsection{RFID Reader}

\begin{figure}[H]
\centering
\includegraphics[width=0.35\textwidth]{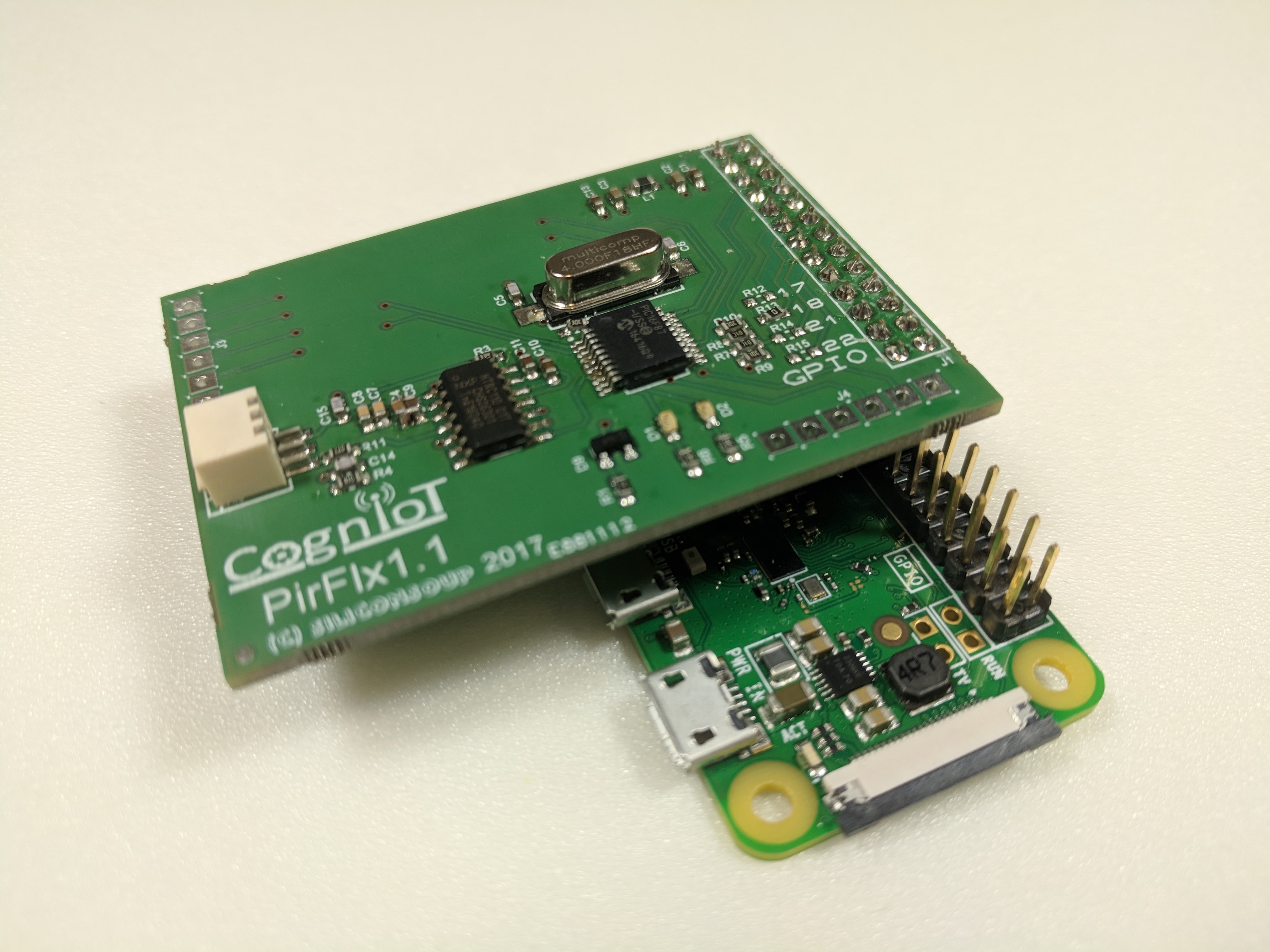}
\caption{The RFID reader attached to the Raspberry Pi.}
\label{computer}
\end{figure}

This system utilizes the 125 kHz RFID reader available from CognIoT\footnote{\url{https://www.tindie.com/products/CognIoT/125khz-rfid-reader-for-raspberry-pi/}}, which connects to GPIO pins 1-36 with the chip overlapping the body of the computer (see Figure \ref{computer}). The reader requires input from an antenna tuned to 770 $\mu$H.

\subsection{Power Supply}

For power we utilized the Poweradd Pilot X7\footnote{\url{https://amzn.to/2OndvHj}}, a 20,000 mAh power bank that can run the setup continuously for four days. These take around 10 hours to charge, so if you require constant data collection it may be a good idea to have two batteries for each birdfeeder that you can cycle through.

\subsection{Tags}

We recommend ordering EM4102 passive integrated transponder (PIT) tags from Eccel Technology, as they come pre-embedded in leg bands of various sizes\footnote{\url{https://eccel.co.uk/product-category/avian-products/bird-tags/}}. Please ensure that you know the band size required for your species before placing an order. We use the 2.6 mm bands for house finches.

\subsection{Antennas}

Building the perch antennas is the most technically challenging aspect of this setup. Each antenna consists of 30 AWG enameled magnet wire\footnote{\url{http://bit.ly/2AQ8xzc}} wrapped around a 40 x 6 mm ferrite rod\footnote{\url{http://bit.ly/2QmIaGw}}. Wrapping the wire around the rod is done by placing one end of the rod in a power drill, anchoring the wire on the other end of the rod with tape, and slowly rotating the rod with the drill. In order to achieve the required inductance of 770 $\mu$H (+/- 50 $\mu$H), we have found that doing four overlapping wraps between 5 mm and 15 mm from one end (range of 10 mm) works well. We recommend having an inductance meter handy for troubleshooting\footnote{\url{http://bit.ly/2DqgT2S}}.

Once wrapping is complete, we cover each end with a 6 mm rubber tip\footnote{\url{https://amzn.to/2DotDHl}} and the entire rod in 8 mm polyolefin heat shrink tubing\footnote{\url{https://amzn.to/2D7Fh8Q}}. Then, the two loose wires can be lightly stripped and soldered to 2 ft long leads\footnote{\url{http://bit.ly/2D83z2b}}. These leads should be attached to the three position 1.5 mm connector\footnote{\url{http://bit.ly/2SRmcNj}} by crimping each wire to the 24-26 AWG terminals\footnote{\url{http://bit.ly/2PFT1ya}} and inserting them into the left-most and right-most positions. The last step is to waterproof the entire antenna using rubber coating spray (i.e. Plasti Dip).

If you have already completed the software and firmware steps below, then you can test the antenna by plugging it into the white connector on the CognIoT board. If it is correctly tuned then the red light should turn green when you present the antenna with a PIT tag. We have found that this antenna design has the highest read range ($\sim$2 cm) on the unwrapped end of the antenna.

\begin{figure}[H]
\centering
\includegraphics[width=0.35\textwidth]{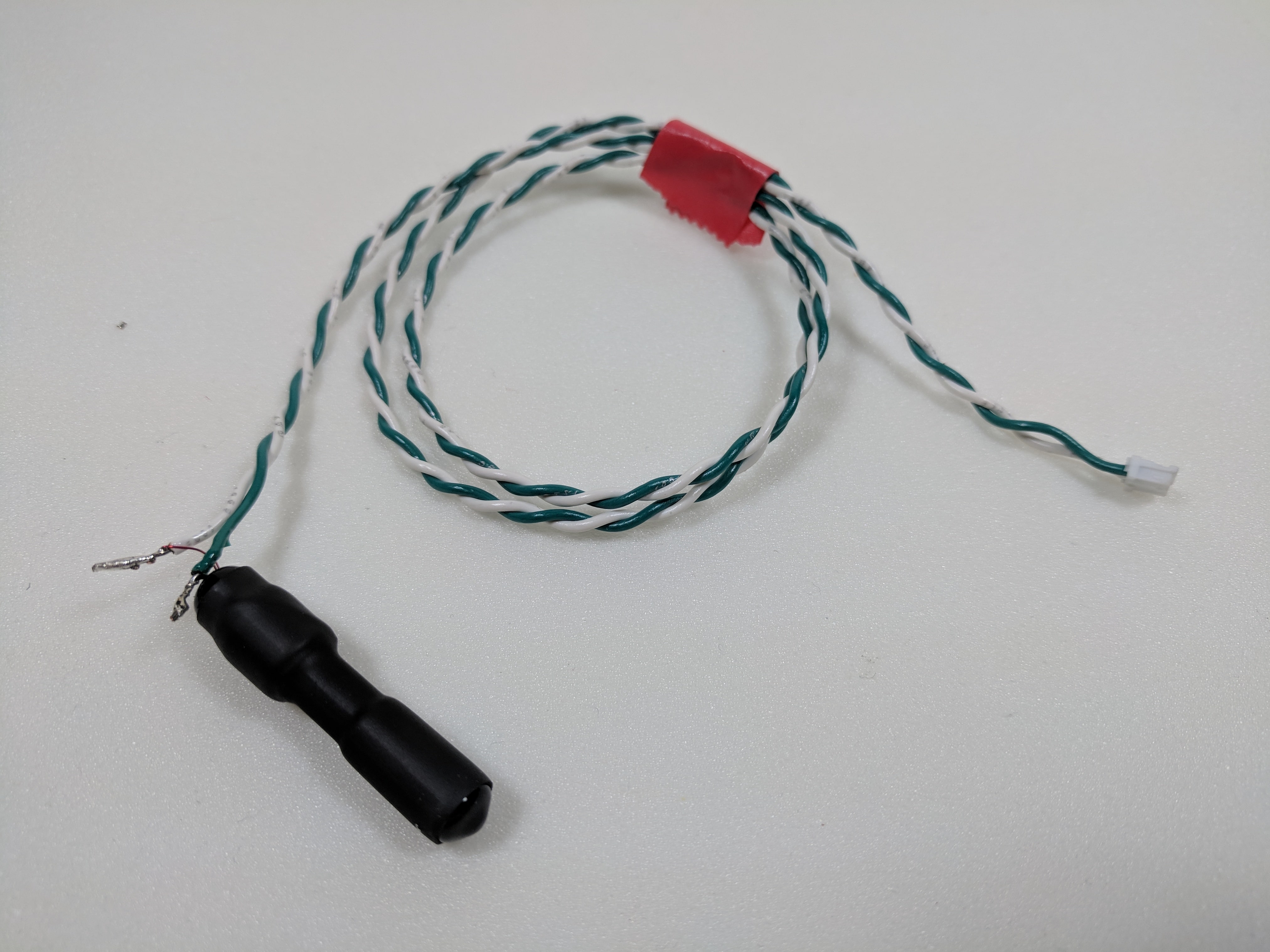}
\caption{The completed antenna.}
\label{antenna}
\end{figure}

\begin{figure}[H]
\centering
\includegraphics[width=0.35\textwidth]{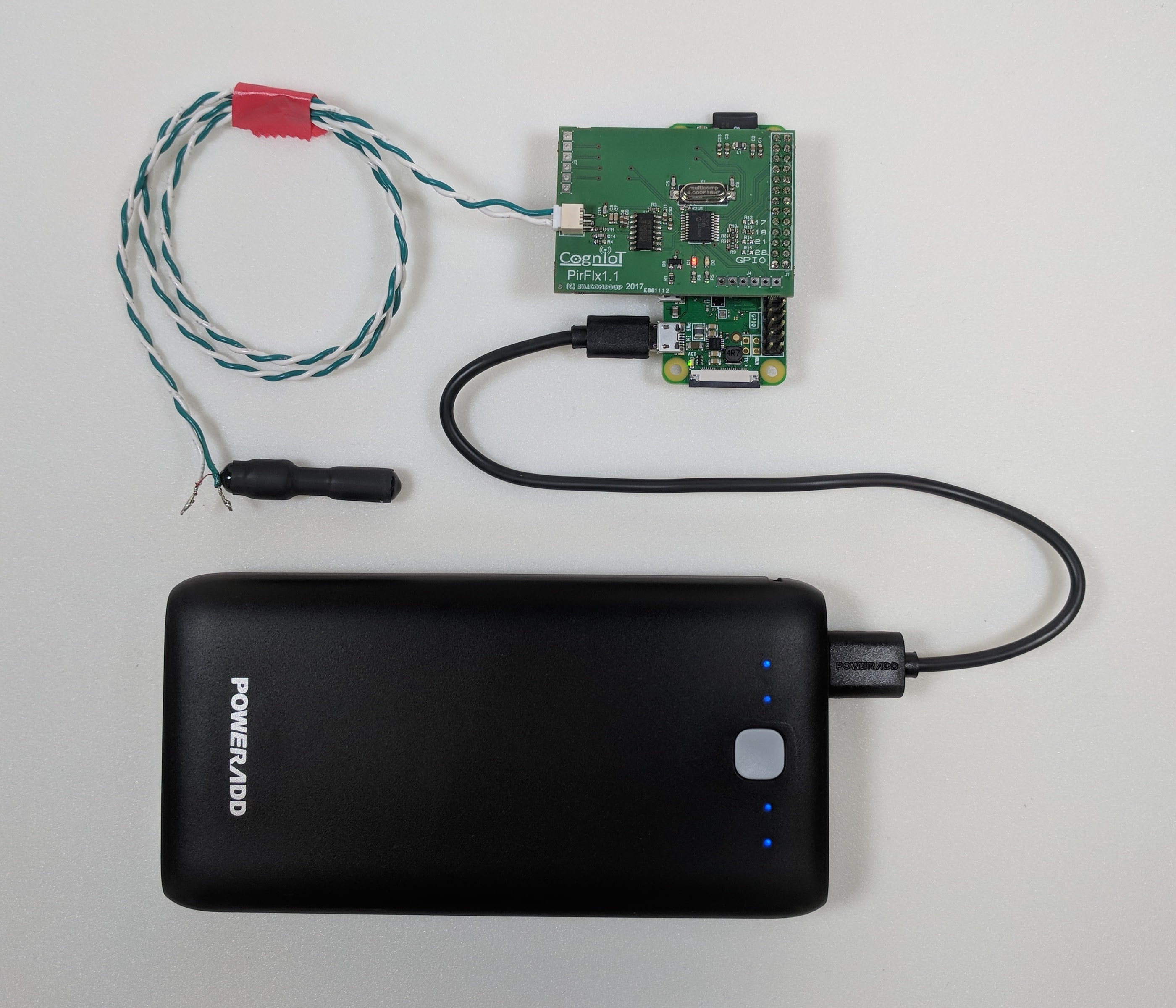}
\caption{The antenna and power supply connected to the computer and RFID reader.}
\label{setup}
\end{figure}

\section*{Software}

The computers in each birdfeeder run a modified Linux operating system with scripts allowing them to continuously collect and store RFID data, and upload it to cloud storage after connecting to a personal hotspot generated by a smartphone. These scripts run automatically when the feeder is connected to a power supply, so no commands need to be run in the field. Following are the steps needed to set up the operating system.

First, download the most recent version of Raspbian Stretch Lite\footnote{\url{https://www.raspberrypi.org/downloads/raspbian/}} and follow the instructions to write the image to a microSD card\footnote{\url{https://www.raspberrypi.org/documentation/installation/installing-images/}}. Once the operating system is installed, boot up the computer and log in using the default username (pi) and password (raspberry). Open the configuration menu:

\begin{lstlisting}[language=bash]
sudo raspi-config
\end{lstlisting}

\noindent Enter ``Network Options'' $\rightarrow$ ``Wi-fi'' to setup the internet connection. Then enter ``Interfacing Options'' $\rightarrow$ ``Serial'', and disable the login shell from using the serial port. Make sure that the serial port hardware is left enabled. Lastly, enable automatic login to the console by entering ``Boot Options''. The computer may require you to reboot before continuing. Next, make sure that the prerequisite software for the RFID reader is installed:

\begin{lstlisting}[language=bash]
sudo apt-get install python-dev python-setuptools python-pip

sudo apt-get install git-core
git clone git://git.drogon.net/wiringPi
cd ~/wiringPi
./build

sudo apt-get install powertop

sudo pip install wiringpi
sudo pip install numpy
\end{lstlisting}

\noindent Install the example software for the RFID reader:

\begin{lstlisting}[language=bash]
git clone https://github.com/CognIot/RFID_125kHz
\end{lstlisting}

\noindent Import ``rfid\_reader.py'' and ``schedule.sh'' from the data repository into /home/pi, using secure copy or file transfer protocol. ``rfid\_reader.py'' is a modified version of a CognIot script\footnote{\url{https://github.com/CognIot/RFID_125kHz}} that interfaces with and controls the RFID reader. If you need to change the default polling delay of the RFID reader (262 ms), edit the SetPollingDelay function in ``rfid\_reader.py'' according to CognIot's documentation. ``schedule.sh'' stores the collected data into files compatible with \textit{feedr}, an R package for managing and visualizing data from RFID-equipped birdfeeders \cite{Lazerte2017}. Make both files executable:

\begin{lstlisting}[language=bash]
sudo chmod +x rfid_reader.py
sudo chmod +x schedule.sh
\end{lstlisting}

\noindent Remember to modify the ``schedule.sh'' script to reflect the name, latitude, and longitude of the site where the birdfeeder will be deployed. 

Next, ``schedule.sh'' needs to run automatically on startup so that a new data file is generated every time the battery is changed. It is also a good idea to run \textit{powertop} to save power. Open the ``rc.local'' file located in /etc. Erase the lines for printing the IP address, and add the following lines above ``exit 0'' so that they run during startup:

\begin{lstlisting}[language=bash]
#use powertop auto-tune to save power
sudo powertop --auto-tune

#run schedule script
/home/pi/schedule.sh &
\end{lstlisting}

\noindent Create a folder in /home/pi for the RFID data:

\begin{lstlisting}[language=bash]
mkrdir rfid_logs
\end{lstlisting}

\noindent If you plan on backing up collected RFID data, install \textit{rclone} and use the standard configuration for your cloud storage service of choice\footnote{\url{https://rclone.org/docs/}}. If you plan on using a cloud storage service that requires web authentication then you need to run the configuration via SSH from a computer with a GUI:

\begin{lstlisting}[language=bash]
curl https://rclone.org/install.sh | sudo bash
rclone config
\end{lstlisting}

\noindent There are several ways to handle backups. If the birdfeeders will be in locations with dependable Wi-Fi connections, then you can set up hourly backups via \textit{cron}. For our situation, we decided to have the birdfeeder automatically upload the RFID data to cloud storage whenever it connects to a personal hotspot generated by a smartphone. The first step is to install and configure an appropriate network manager:

\begin{lstlisting}[language=bash]
sudo apt-get install wicd wicd-curses

sudo systemctl stop dhcpcd
sudo systemctl disable dhcpcd
sudo systemctl start wicd.service
sudo systemctl enable wicd.service
sudo gpasswd -a pi netdev
\end{lstlisting}

\noindent Next, open up the text-based interface for the manager to prioritize and automatically connect to the personal hotspot, and disable automatic connection to any networks that might interfere (e.g. institutional connections requiring login):

\begin{lstlisting}[language=bash]
wicd-curses
\end{lstlisting}

\noindent Import ``drive\_backup.sh'' from the data repository into /etc/wicd/scripts/postconnect using secure copy or file transfer protocol, remove the ``.sh''file extension, and make it executable:

\begin{lstlisting}[language=bash]
sudo chmod +x /etc/wicd/scripts/postconnect/drive_backup
\end{lstlisting}

\noindent ``drive\_backup'' uploads everything in /home/pi/rfid\_logs to /rfid\_logs/site\_name, a directory that you need to create in your cloud storage. Remember to change ``site\_name'' to the name of the site where the birdfeeder will be deployed.

Lastly, before the RFID reader can be used with the computer it must be set up to read EM4102 tags. Until this step is completed the reader will not register correctly constructed antennas. Run the setup script and follow the relevant menu options:

\begin{lstlisting}[language=bash]
python RFID_125kHz/python/RFIDReader.py
\end{lstlisting}

\section*{Example}

In the spring and summer of 2018, we banded 138 house finches with PIT tags (immature: 92.0\%; adult female: 3.6\%; adult male: 4.4\%). Due to high levels of dispersal ($\sim$91.3\% based on return rate), we banded an additional 47 house finches in the spring and summer of 2019 (immature: 63.8\%; adult female: 8.5\%; adult male: 27.7\%). All birds were captured using perch traps\footnote{\url{http://www.thirdwheel.biz/perch-traps.html}}, and adult female and immature birds were distinguished based on the color of their secondary covert feathers \cite{Hill2002}.

\begin{figure}[H]
	\centering
	\includegraphics[width=1\textwidth]{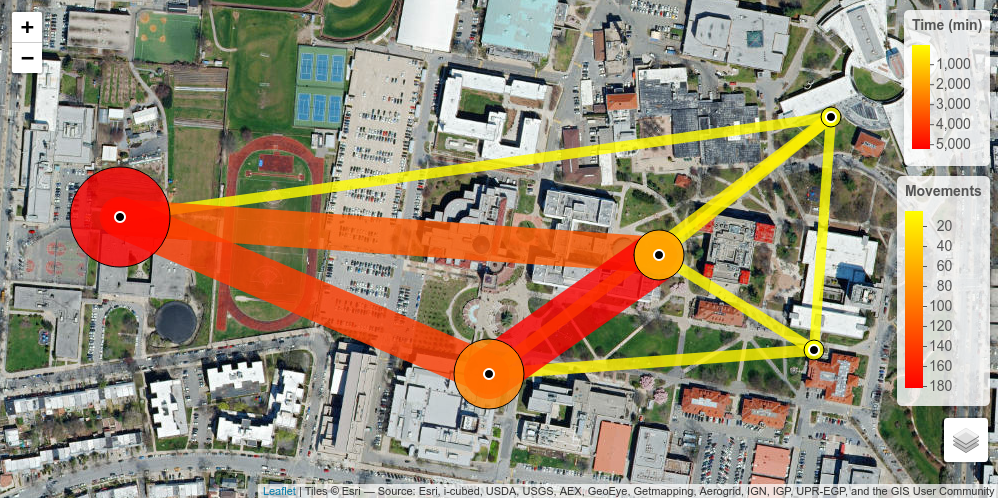}
	\caption{The locations of the five birdfeeders on the Queens College campus, the total number of minutes birds spent at each location, and the total number of times birds moved between locations (disregarding the amount of time between visits). Plotted using \textit{feedr} \cite{Lazerte2017}.}
	\label{map}
\end{figure}

\begin{figure}[H]
	\centering
	\includegraphics[width=1\textwidth]{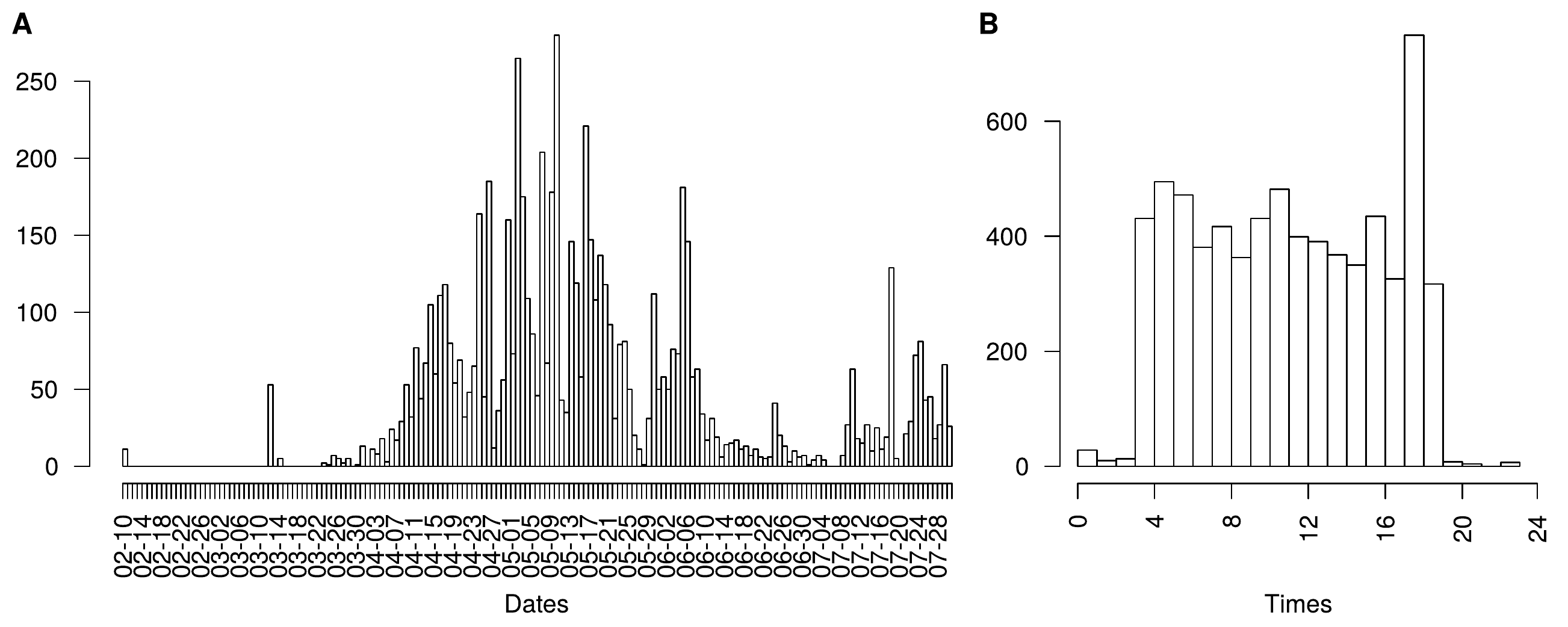}
	\caption{(A) The number of feeder visits (\textit{y}-axis) that occurred on each day of the study period. (B) The total number of feeder visits (\textit{y}-axis) that occurred during each hour across the entire study period.}
	\label{datetime}
\end{figure}

Five RFID-equipped birdfeeders were deployed on the Queens College campus from October 31, 2018 to August 1, 2019. Every four days, we visited each birdfeeder to replace the food and batteries and reboot them with a personal hotspot in range. Over the course of the field season we collected 6,878 visits from 28 individuals (12 banded in 2018, and 16 banded in 2019). The total time in minutes that individuals spent at each of the five feeders, as well as the number of movements between feeders, can be seen in Figure \ref{map}. Birds began to visit the feeders regularly towards the end of March, and activity peaked in early May (left panel of Figure \ref{datetime}). Foraging activity was relatively constant throughout the day with the highest levels occurring around dusk (right panel of Figure \ref{datetime}), consistent with the findings of Bonter et al. \cite{Bonter2013}.

Once the raw visitation data has been imported into R using the package \textit{feedr} \cite{Lazerte2017}, it can be used for several different analyses. Here we demonstrate two examples: (1) the estimation of the social network of the population, and (2) the estimation of the dominance hierarchy of the population.

\begin{figure}[H]
	\centering
	\includegraphics[width=0.8\textwidth]{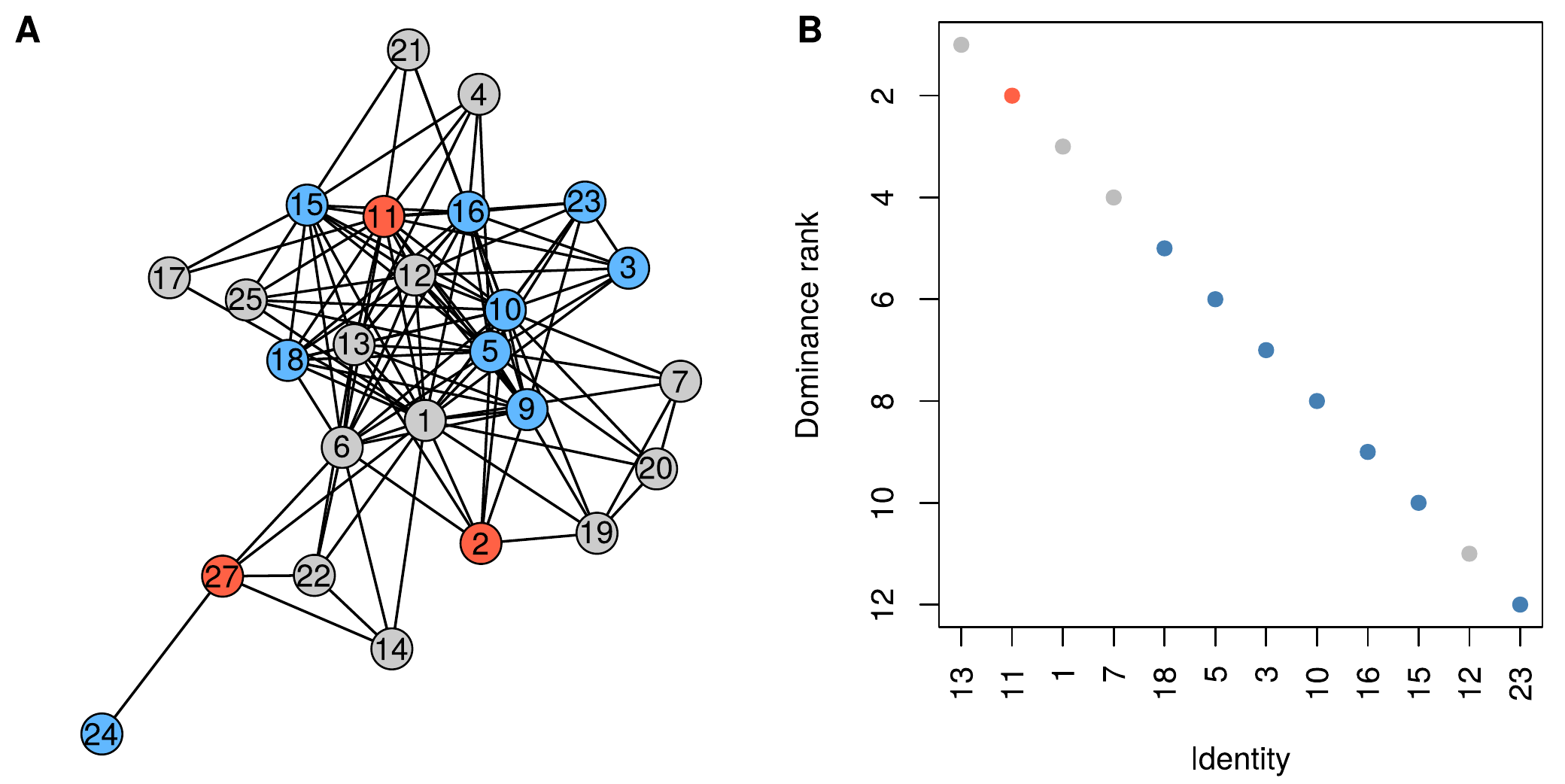}
	\caption{(A) The estimated social network of the 25 birds for which we had association data. (B) The dominance ranks (\textit{y}-axis) of the 12 birds for which we had displacement interactions. Plotted with \textit{aniDom} \cite{Farine2019}.}
	\label{netdom}
\end{figure}

For species such as house finches that congregate around birdfeeders, co-occurrences of individuals during foraging bouts can be used to estimate the social network structure of a population \cite{Adelman2015}. Firstly, adjacent feeder visits by different birds are transformed into association data using machine learning with a Gaussian mixture model \cite{Psorakis2015}, implemented in the R package \textit{asnipe} \cite{Farine2013}. This association data is then used to reconstruct a weighted social network of the population, where nodes are individual birds and links represent interaction rates between them. The estimated social network of the population can be seen in the left panel of Figure \ref{netdom}. In this case, the 25 birds represent a subsample of the entire population. Although partial networks often closely reflect the structure of full networks \cite{Silk2015}, any social network constructed from a subsample should be interpreted with caution.

House finches have linear dominance hierarchies in which females are typically dominant to males \cite{Belthoff1991,Thompson1960a}, and aggressive interactions between individuals often occur at food sources when one individual displaces another \cite{Thompson1960b}. As such, investigating dominance using displacement data is a well-established method in house finches \cite{McGraw2002,Moyers2018} and other species \cite{Miller2017,Evans2018}. Firstly, displacement events, in which one bird displaces another at the feeder within two seconds, are extracted from the raw visitation data. This pairwise interaction data is then analyzed using the \textit{aniDom} package in R \cite{Farine2019}. Birds who consistently displace other individuals at the feeders are coded as dominant to those individuals. The estimated dominance hierarchy of the population can be seen in the right panel of Figure \ref{netdom}.

During the field season we periodically experienced three minor technical issues. Firstly, cold weather and high humidity can cause the batteries to shut down prematurely. If temperatures are approaching freezing or rain is expected, it is a good idea to check the batteries daily rather than every four days. Next, sometimes the configuration file for \textit{rclone} is cleared when the backup script runs without the personal hotspot. If files are not appearing in the cloud backup, a quick reconfiguration of \textit{rclone} should solve the issue. Lastly, if the computers reboot without connecting to a personal hotspot the system time will not be accurate. Be sure that the personal hotspot is on and within range whenever the computers are rebooted.

\section*{\large Acknowledgments}

I would like to thank Alec Lindsay and Andrew Richards for their invaluable help with antenna design, as well as Shari Zimmerman and Andrea Lopez for their assistance with birdfeeder maintenance and data collection. I would also like to thank David Lahti, as well as all members of the Lahti lab, for their valuable conceptual and analytical feedback.

\section*{\large Data Availability Statement}

All required scripts are available in the Harvard Dataverse repository: \url{https://doi.org/10.7910/DVN/XAIRNM}.

\renewcommand*{\bibfont}{\scriptsize}
\printbibliography[title=\large References]

\end{document}